\documentclass[sn-mathphys,Numbered]{sn-jnl}
\usepackage{graphicx}%
\usepackage{multirow}%
\usepackage{amsmath,amssymb,amsfonts}%
\usepackage{amsthm}%
\usepackage{mathrsfs}%
\usepackage[title]{appendix}%
\usepackage{xcolor}%
\usepackage{textcomp}%
\usepackage{manyfoot}%
\usepackage{booktabs}%
\usepackage{subfigure}
\usepackage{algorithm}%
\usepackage{algorithmicx}%
\usepackage{algpseudocode}%
\usepackage{listings}%
\theoremstyle{thmstyleone}%

\theoremstyle{thmstyletwo}%

\theoremstyle{thmstylethree}%

\begin{document}

\title[A New Generalized Fisk distribution:]
{A New Generalized Fisk distribution:}
\subtitle{Its Properties, Characterizations and Applications}

\author*[1]{\fnm{Veeranna} \sur{Banoth}}\email{veerusukya40@gmail.com}

\affil*[1]{\orgdiv{DST-Center for Interdisciplinary Mathematical Science, Institute of Sciences}, \orgname{Banaras Hindu University}, \orgaddress{\city{Varanasi}, \state{Uttar Pradesh}, \country{India}, \postcode{221005}}}


\abstract{The shortcomings of the traditional univariate distributions in the past greatly encouraged mathematical statisticians to develop new generalizations of distributions. The New Generalized Fisk distribution, a unique distribution presented in this study, is thoroughly examined. There is a thorough discussion of a few fundamental statistical traits and attributes, such as the quantile function, order statistics, skewness and kurtosis, moments, and moment-generating functions. The new distribution incorporates additional parameters, enhancing its ability to capture a wide range of skewness and kurtosis behaviors, making it applicable to diverse fields such as economics, reliability engineering, and environmental sciences. Both numerical and graphical evaluations are used to evaluate the performance of the recently proposed model. Additionally, the performance of the maximum likelihood estimators is assessed by a simulation study. Real-world applications are analyzed, and the model parameters are estimated using the maximum likelihood estimation technique. It is contrasted with the popular models of computing. According to model adequacy and discrimination approaches, the suggested model performs the best. The models are compared to choose the model that best fits the data with the essential characteristics. The graphical and model comparison approaches suggested an outstanding improvement in the combined distribution.}


\maketitle

\section{Introduction}\label{sec1}
A lot of applied statisticians are interested in the theory of probability distribution. The competence in applied statistics has also been interested in the creation and introduction of new families of distributions. Several strategies are frequently employed, including the extension, modification, adaption, and application of extra parameters to the current models.

Some distributions are naturally flexible, both in theory and in practice.  This serves as the primary motivator for the area's increased attention.
The odd generalized half logistic Weibull-G family of distributions is a new generalized family of distributions that was recently proposed by Chipepa et al. (2020).  In order to talk about certain structural characteristics of the newly suggested family of distributions using the failure time data, they additionally took into account a few unique baseline distributions.  They have demonstrated that this model outperforms its rival models. 

A new extended flexible Weibull distribution (NEx-FW), which gives more flexibility, was proposed by Ahmad and Hussain (2017) to simulate lifetime data with bathtub-shaped failure rates.  They have also brought forward a class of generalizations of the Weibull model that have been put up by numerous academics in the literature.  Examples include the generalized modified Weibull (GMW) distribution put forth by Carrasco et al. (2008), the exponentiated modified Weibull extension (EMWEx) distribution presented by Sarhan and Apaloo (2013), the flexible Weibull (FWEx) distribution of Bebbington et al. (2007), the Beta-Weibull (BW) distribution of Famoye et al. (2005), and the Kumaraswamy Weibull (KW) distribution proposed by Cordeiro et al. (2010).

In this paper, we suggest a novel generalized-X (NG-X) family of distributions that can be applied to any baseline distribution in the fields of lifespan research, biomedical engineering, health, survival analysis, and reliability.  By using a log logistic distribution as the baseline distribution and referring to it as NG-Fisk, we demonstrated the viability of the NG-X family. The majority of NG-Fisk's statistical and practical characteristics are covered, and a numerical analysis is carried out using simulation and actual data.

The remaining sections of the essay are summarized as follows:  Section 2 introduces a novel class of distributions and related computations and graphical illustration of a specific member of the proposed distribution family. While, Section 3 provides a detailed statistical properties of a proposed model.   Section 4 covers basic statistical Inferences of estimation techniques of the distribution family. Section 5 covers the simulation studies, while Section 6 compares models and applies the distribution to real data. Section 7 concludes with a recommendation.

\section{A New Generalized Fisk distribution}\label{sec2}
Let the probability density function $f(x; \alpha, \beta)$ of the Fisk distribution (is also called log-logistic distribution), its p.d.f. is given by $f(x; \alpha, \beta) = \frac{(\frac{\beta}{\alpha})(\frac{x}{\alpha})^{(\beta-1)}}{(1+(\frac{x}{\alpha})^\beta)^2}$ and the cumulative distribution function (C.D.F.) $F(x; \alpha, \beta)$ of the Fisk distribution is given by $F(x; \alpha, \beta) = \frac{1}{1+(\frac{x}{\alpha})^{-\beta}}$:\\
Now, we computed the new generalized-Fisk distribution of the probability density function and its a cumulative distribution function as follows:
\begin{equation}\label{pdf1}
\begin{split}
	g(x;\theta, \delta,\underline{\omega}) &= \frac{\theta(1-\delta)f(x; \alpha, \beta)[1-F(x;\alpha,\beta)]^{\theta-1}}{[1-\delta F(x;\alpha,\beta)]^{\theta+1}}\\
	&=\frac{\theta(1-\delta)[\frac{(\frac{\beta}{\alpha})(\frac{x}{\alpha})^{(\beta-1)}}{(1+(\frac{x}{\alpha})^\beta)^2}][1-\frac{1}{1+(\frac{x}{\alpha})^{-\beta}}]^{\theta-1}}{[1-\delta (\frac{1}{1+(\frac{x}{\alpha})^{-\beta}})]^{\theta+1}}\\
	g(x;\theta, \delta,\underline{\omega}) &= \frac{\theta(1-\delta)(\frac{\beta}{\alpha})(\frac{x}{\alpha})^{\beta-1}}{[1+(1-\delta)(\frac{x}{\alpha})^\beta]^{\theta+1}}
	\end{split}
\end{equation}
\begin{equation}
	\begin{split}
	G(x;\theta, \delta,\underline{\omega}) &= 1-(\frac{\overline{F}(x;\underline{\omega})}{1-\delta F(x;\underline{\omega})})^\theta\\
	&= 1-\big(\frac{1-\frac{(\frac{x}{\alpha})^\beta}{1+(\frac{x}{\alpha})^\beta}}{1-\delta (\frac{(\frac{x}{\alpha})^\beta}{1+(\frac{x}{\alpha})^\beta})}\big)^\theta\\
	G(x;\theta, \delta,\underline{\omega}) &= 1- (\frac{1}{1+(1-\delta)(\frac{x}{\alpha})^\beta})^\theta; ~~~~~~x>0,
	\end{split}
\end{equation}
respectively.\\
Where $\underline{\omega} = (\alpha,\beta); \alpha>0, \beta> 0, \theta>0$ and $\delta \in (0,1)$.\\
The Survival function, hazard rate function, cumulative hazard rate function, and reverse failure rate function for the new generalized Fisk distribution are given by 
\begin{equation}
\begin{split}
S(X) &= \big(1-\frac{(1-\delta)(F(X))}{1-\delta F(X)}\big)^\theta \\
&= \Bigg[1-\big(\frac{(1-\delta)(\frac{(\frac{x}{\alpha})^\beta}{1+(\frac{x}{\alpha})^\beta})}{1-\delta(\frac{(\frac{x}{\alpha})^\beta}{1+(\frac{x}{\alpha})^\beta})}\big) \Bigg]^\theta \\
S(X) &= \big[\frac{1}{1+(1-\delta)(\frac{x}{\alpha})^\beta}\big]^\theta,
\end{split}
\end{equation}
\begin{equation}
	\begin{split}
	h(X) &= \frac{\theta(1-\delta)f(X)}{(1-F(X))(1-\delta F(X))} \\
	&= \frac{\theta(1-\delta)\frac{(\frac{\beta}{\alpha})(\frac{x}{\alpha})^{(\beta-1)}}{(1+(\frac{x}{\alpha})^\beta)^2}}{(1-\frac{1}{1+(\frac{x}{\alpha})^{-\beta}})(1-\delta(\frac{1}{1+(\frac{x}{\alpha})^{-\beta}}))} \\
	h(X) &= \frac{\theta(1-\delta)(\frac{\beta}{\alpha})(\frac{x}{\alpha})^{\beta-1}}{1+(1-\delta)(\frac{x}{\alpha})^\beta},
	\end{split}
	\end{equation}
	
\begin{equation}
\begin{split}
H(X) &= -log\Bigg\{1-[\frac{(1-\delta)F(X)}{1-\delta F(X)}]\Bigg\}^\theta \\
H(X) &= -log\big\{\frac{1}{1+(1-\delta)(\frac{x}{\alpha})^\beta}\big\}^\theta ,
\end{split}
\end{equation}	
and
\begin{equation}
\begin{split}
r(X) &= \frac{\theta(1-\delta)f(X)(1-F(X))^{\theta-1}}{(1-\delta F(X))([1-\delta F(X)]^\theta-[1-F(X)]^\theta)} \\
&=\frac{\theta(1-\delta)(\frac{(\frac{\beta}{\alpha})(\frac{x}{\alpha})^{(\beta-1)}}{(1+(\frac{x}{\alpha})^\beta)^2})(1-\frac{1}{1+(\frac{x}{\alpha})^{-\beta}})^{\theta-1}}{(1-\delta (\frac{1}{1+(\frac{x}{\alpha})^{-\beta}}))([1-\delta (\frac{1}{1+(\frac{x}{\alpha})^{-\beta}})]^\theta-[1-(\frac{1}{1+(\frac{x}{\alpha})^{-\beta}})]^\theta)} \\
r(X) &= \frac{\theta(1-\delta)(\frac{\beta}{\alpha})(\frac{x}{\alpha})^{\beta-1}(1+(\frac{x}{\alpha})^\beta)^{-\theta}}{(1+(1-\delta)(\frac{x}{\alpha})^\beta)((\frac{1+(1-\delta)(\frac{x}{\alpha})^\beta}{1+(\frac{x}{\alpha})^\beta})^\theta-(\frac{1}{1+(\frac{x}{\alpha})^\beta})^\theta)},
\end{split}
\end{equation}
respectively.\\
Figure 1 shows a graphical representation of the CDF, PDF and hazard function of a novel generalized Fisk distribution.  In Figure 1, the graphs in parts (a), (b) and (c) illustrate the CDF, PDF and the hazard rate function of a novel generalized Fisk distribution, respectively, to demonstrate varied patterns and the distribution's flexibility.  The PDF (as defined by equation (1)) is right and left skewed, symmetric, and decreasing.  Furthermore, the hazard rate function exhibits dropping, increasing, and parabolic up trends. Similarly, the depending on the parameters, the distribution exhibits early steep rises or gradual climbs, reflecting left-skewed, right-skewed, or symmetric behavior.

\begin{figure}
	\centering
	\subfigure[]{\includegraphics[width=0.8\textwidth]{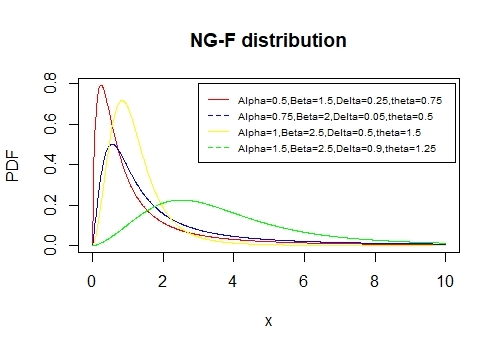}} 
	\subfigure[]{\includegraphics[width=0.8\textwidth]{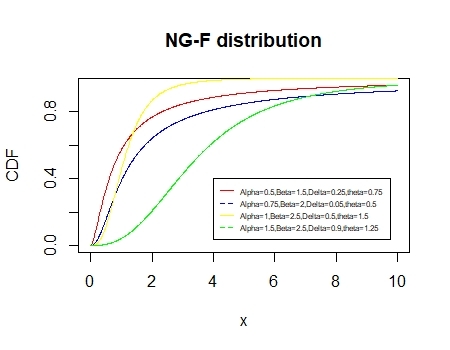}} 
	\subfigure[]{\includegraphics[width=0.8\textwidth]{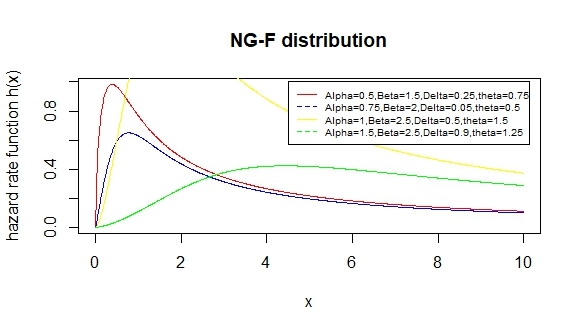}}

	\caption{(a) PDF plot (b) CDF plot (c) Hazard rate function plot for the NG-Fisk distribution }
	\label{fig:foobar}
\end{figure}

\newpage
\section{Statistical properties of a new generalized-Fisk distribution} \label{sec:Statistical properties}
In this section, some of the important statistical properties of the new proposed family of distributions are discussed.

\subsection{Quantile function}
Quantile function can be used for many purposes in theory and numerical applications in statistics. For example, it can be used to draw simulations. We can obtain the the quantile function for the a new generalized-Fisk distributions by applying the inverse technique. Thus,
\begin{equation*}
Qx_n(p) = G(x; \theta, \delta, \alpha, \beta)^{-1},
\end{equation*}
\begin{equation*}
1-(1-[\frac{(1-\delta)F(x;\alpha,\beta)}{1-\delta F(x;\alpha,\beta)}])^\theta=p, ~~~~~~0<p<1.
\end{equation*}
By using a simple algebraic application and solving the non-linear equation, this can be computed as follows:
\begin{equation}
Qx_n(p) = F^{-1}(\frac{1-(1-p)^{1/\theta}}{-\delta(1-p)^{(1/\theta) +1}}),
\end{equation} 
where $p \sim Uniform(0,1)$.\\
The median(Med) can be obtained by substituting p=1/2 in equation (7) as:
\begin{equation*}
Med = F^{-1}[\frac{1-(1-1/2)^{1/\theta}}{-\delta(1-1/2)^{(1/\theta) +1}}],
\end{equation*}
which can be simplified to
\begin{equation*}
Med = F^{-1}[\frac{(1-(1/2)^{1/\theta})}{-\delta(1/2)^{\frac{1}{\theta}+1}}].
\end{equation*}
Similarly, one can obtain the lower and upper quantiles by substituting p=1/4 and p=3/4 in equation (7), respectively.
\subsection{Skewness and Kurtosis:}
 Based on the formulas given by Galton (1883) and Moors (1988), we can obtain skewness and kurtosis(K) from equation (7) as
 \begin{equation*}
 S_k = \frac{(Q_3-2Q_2+Q_1)}{(Q_3-Q_1)} = \frac{(q_{(0.75)}-2 q_{(0.5)}+q_{(0.25)})}{(q_{(0.75)}-q_{(0.25)})},
 \end{equation*}
 and
 \begin{equation*}
 K = \frac{(Q_{7/8}-Q_{5/8}+Q_{3/8}-Q_{1/8})}{(Q_{6/8}-Q_{2/8})} = \frac{(q_{(-.875)}-q_{(0.625)}+q_{(0.375)}-q_{(0.125)})}{(q_{(0.75)}-q_{(0.25)})},
 \end{equation*}
 respectively.
 \subsection{Order Statistics:}
 It is well-common that order statistics are widely used in the statistical applications, such as reliability and lifetime testing. Suppose that $X_1, X_2, ... , X_n$ are a random sample of size $n$ drawn independently from the a new generalized-Fisk family of distributions with parameters $\theta, \delta, \alpha, and \beta$. Let $X_{1:n}, X_{2:n}, ..., X_{n:n}$ be the corresponding order statistics. Then, the density of $X_{i:n}$ for $(i=1,2,..,n)$ is given by 
 \begin{equation*}
 f_{(i:n)}(x) = \frac{n!}{(i-1)!(n-i)!}\sum_{j=0}^{n-i}(-i)^j\binom{n-i}{j}f(x;\alpha,\beta)[F(x;\alpha,\beta)]^{i+j-1},
 \end{equation*}
 \begin{equation*}
 \begin{split}
f_{(i:n)}(x) &= \frac{n!}{(i-1)!(n-i)!}\sum_{j=0}^{n-i}(-i)^j\binom{n-i}{j}\frac{\theta(1-\delta)f(x;\alpha,\beta)[1-F(x;\alpha,\beta)]^{\theta-1}}{[1-\delta(1-F(x;\alpha,\beta))]^{\theta+1}} \\
&* [1-[\frac{1-F(x;\alpha,\beta)}{1-\delta F(x;\alpha,\beta)}]^\theta]^{i+j-1}.
 \end{split}
 \end{equation*}
 The order statistics of the special member of the family a new generalized-Fisk distribution is derived in the same way as follows:
 \begin{equation*}
 \begin{split}
 f_{(i:n)G-F}(x;\alpha,\beta,\theta,\delta) &= \frac{n!}{(i-1)!(n-i)!}\sum_{j=0}^{n-i}(-i)^j\binom{n-i}{j}\frac{\theta(1-\delta)(\frac{\beta}{\alpha})(\frac{x}{\alpha})^{\beta-1}}{(1+(1-\delta)(\frac{x}{\alpha})^\beta)^{\theta+1}} \\
 &*[[\frac{1}{1+(1-\delta)(\frac{x}{\alpha})^\beta}]^\theta]^{i+j-1}.
 \end{split}
 \end{equation*}
 This can be simplified in equation (8) as:
 \begin{equation}
f_{(i:n)G-F}(x;\alpha,\beta,\theta,\delta) = \sum_{j=0}^{n-i}\eta_j f_{G-F}(x;\alpha,\beta,\theta,\delta),
 \end{equation}
 where $\eta_j = \frac{n!}{(i-1)!(n-i)!}\sum_{j=0}^{n-i}(-i)^j\binom{n-i}{j} $.
 \subsection{Moments and Moment Generating Functions:}
 The $r^{th}$ central moment of the new proposed family can be obtained as follows:
 \begin{equation*}
 \begin{split}
 \mu_{r}^{'} &= \int_{-\infty}^{+\infty}x^rg(x;\alpha,\beta,\theta,\delta)dx \\
 &= \int_{-\infty}^{+\infty}x^r \frac{[\theta(1-\delta)]^{i+j}f(x;\alpha,\beta)[1-F(x;\alpha,\beta)]^{\theta i-1}}{[1-\delta F(x;\alpha,\beta)]^{\theta i+1}}dx \\
 &= \sum_{i=0}^{\infty}\frac{[\theta(1-\delta)]^{i+1}}{i!}\eta_{r,i+1}.
 \end{split}
 \end{equation*}
 Moment generating function, $M_x(t)$ for the family is given in equation (9) as:
 \begin{equation}
 M_x(t) = \sum_{i=0}^{\infty}t^r \frac{[\theta(1-\delta)]^{i+1}}{r! i!}\eta_{r,i+1},
 \end{equation}
 where $\eta_{r,i+1} = \int_{-\infty}^{+\infty}x^r \frac{f(x;\alpha,\beta)[1-F(x;\alpha,\beta)]^{\theta i-1}}{[1-\delta F(x;\alpha,\beta)]^{\theta i+1}}dx$.
 
 \section{Statistical Inferences:}
 \subsection{Maximum Likelihood Estimation:}
 In this section, the maximum likelihood estimators (MLE's) for the model parameters of a new generalized-Fisk distribution is dealt in this section. Let $x_1, x_2, ... , x_n$ be observed values of a sample randomly selected from a new generalized-Fisk distribution with parameters $\theta, \delta, \alpha, \beta$. Given the PDF of the new proposed family of distribution in equation (1) and the total likelihood function (in equation (10) below)
 \begin{eqnarray}
 L(x;\alpha,\beta,\theta,\delta) = \prod_{i=1}^{n}\frac{\theta(1-\delta)f(x;\alpha,\beta)[1-F(x;\alpha,\beta)]^{\theta-1}}{[1-\delta F(x;\alpha,\beta)]^{\theta+1}},
 \end{eqnarray}
 the log-likelihood function of the respective sample $logL(x;\Theta)$ is given below in equation (11) and the model parameters can be estimated by taking the first partial derivative of the $logL(x;\Theta)$ with respect to each model parameters and equating to zero and solving them simultaneously. 
 \begin{equation}
\begin{split}
 logL(x;\Theta) &= n log \theta + n log(1-\delta)+ \sum_{i=1}^{n}log f(x_i;\alpha,\beta)+ (\theta-1)\sum_{i=1}^{n}log[1-F(x_i;\alpha,\beta)]\\
 &-(\theta+1)\sum_{i=1}^{n}log(1-\delta F(x_i;\alpha,\beta)).
 \end{split}
 \end{equation}
 The partial derivatives of $logL(x;\Theta)$ with respect to each parameters are given below
 \begin{eqnarray}
 \frac{\partial{logL}}{\partial{\theta}} = \frac{n}{\theta} + \sum_{i=1}^{n}log(1-F(x_i;\alpha,\beta)) - \sum_{i=1}^{n}log(1-\delta F(x_i;\alpha,\beta)),
 \end{eqnarray} 
 \begin{eqnarray}
\frac{\partial{logL}}{\partial{\delta}} = -\frac{n}{1-\delta}-(\theta+1)\sum_{i=1}^{n}[\frac{\partial{log(1-\delta F(x_i;\alpha,\beta))}}{\partial{\delta}}],
 \end{eqnarray}
 and
 \begin{eqnarray}
\frac{\partial{logL}}{\partial{\underline{\omega}}} = \sum_{i=1}^{n}\frac{\partial{logf(x_i;\underline{\omega})}}{\partial{\underline{\omega}}} + \frac{(\theta-1)\sum_{i=1}^{n}\partial{log[1-f(x_i;\underline{\omega})]}}{\partial{\underline{\omega}}}-\frac{(\theta+1)\sum_{i=1}^{n}\partial{log(1-\delta F(x_i;\underline{\omega}))}}{\partial{\underline{\omega}}}.
 \end{eqnarray}
 Now, the MLE's of the parameters $\theta,\delta,\underline{\omega}= (\alpha,\beta)$ can be obtained by solving the non-linear equation
 \begin{equation*}
 u_n = \Big(\frac{\partial{logL(x;\Theta)}}{\partial{\theta}}, \frac{\partial{logL(x;\Theta)}}{\partial{\delta}}, \frac{\partial{logL(x;\Theta)}}{\partial{\underline{\omega}}}\Big)^T = 0,
 \end{equation*}
using a numerical method.

\section{Simulation Study}
The performance of the MLEs for a fixed sample size $n$ is assessed in this section.  The performance of the MLEs is investigated numerically for a new Generalized-Fisk model, which is a specific case of the suggested family.  The R programming language was used to calculate the estimators, biases, and practical mean square errors (MSEs).
The following is an itemized list of the empirical stages.\\
(i) For the purpose of calculating the aforementioned quantities, a series of random samples $x_1,x_2,..., x_n$ of sizes $n=25,50, 100,150,250$, and $500$ are collected.  The inverse approach is used to create the samples from a new Generalized-Fisk distribution.\\
(ii) In order to iteratively assess the MLEs for each parameter and sample size, three scenarios are taken into consideration for the four parameters of the suggested model. In case 1, these are grouped by case: $\alpha=1.5, \beta=2.0,\theta=2.5, \delta=0.25$; in case 2, they are $\alpha=1, \beta=3.0,\theta=2.5, \delta=0.5$; and in case 3, they are $\alpha=1.5, \beta=3.5,\theta=2.0, \delta=0.75$.\\
(iii) To calculate the bias and MSEs for each parameter, these samples were run 1000 times.\\
(iv) It provides the mean and variance of the parameters together with the computational formulas for bias and MSEs.
\begin{equation*}
\hat{\eta} = \frac{1}{1000}\sum_{i=1}^{1000}\eta_i, ~~~~ Bias(\hat{\eta}) = \hat{\eta_i}-\eta_i,
\end{equation*}
\begin{equation*}
Var(\hat{\eta}) = \frac{1}{1000}\sum_{i=1}^{1000}(\eta_i-\eta)^2,
\end{equation*}
and
\begin{equation*}
MSE(\hat{\eta}) = Var(\hat{\eta})+(Bias(\hat{\eta}))^2
\end{equation*}
where $\hat{\eta} = (\hat{\alpha}, \hat{\beta}, \hat{\theta}, \hat{\delta})$.\\
Table 1,2, and 3 shows the numerical results for the MLE's, MSE's, bias and their 95\% C.I.  As the sample size increases, the estimated parameter values become more consistent and closer to the genuine parameter values.  And there is a clear indication that as the sample size grows, the error decreases as expected.  Table 1, 2, and 3 shows the findings of the simulation study for several simulated scenarios. 

 \begin{table}[ht]
	\centering
	\caption{Case-1: Simulation study for different parameters values and 95\% C.I.}
	\normalsize
		\begin{tabular}{ |c|c|c |c| c| c|} 
			\hline
			Sample Size & Parameters & MLE & MSEs(MLE) & Bias(MLE)  & 95\% C.I.  \\
			\hline
			\multirow{4}{1em}{25} & $\alpha=1.5$ & 1.7799 & 0.6609 & 0.2799 & (0.593, 3.220) \\
			& $\beta=2.0$ & 2.3334& 1.1447 & 0.3334 & (1.391, 5.091)  \\ 
			& $\theta=2.5$ & 5.5008 & 25.3549 & 3.0008 & (0.447, 10.00)  \\ 
			& $\delta=0.25$ & 0.3367 & 0.0930 & 0.0867 & (0.010, 0.846) \\
			\hline
			\multirow{4}{1em}{50} & $\alpha=1.5$ & 1.6820 & 0.4024 & 0.1820 & (0.754, 3.106) \\
			& $\beta=2.0$ & 2.1207& 0.2444 & 0.1207 & (1.490, 3.236)  \\ 
			& $\theta=2.5$ & 4.5961 & 16.9059 & 2.0961 & (0.708, 10.00)  \\ 
			& $\delta=0.25$ & 0.3020 & 0.0867 & 0.0520 & (0.010, 0.848)  \\
			\hline
			\multirow{4}{1em}{100} & $\alpha=1.5$ & 1.6620 & 0.1817 & 0.1220 & (0.963, 2.697) \\
			& $\beta=2.0$ & 2.0500 & 0.0820 & 0.0500 & (1.597, 2.680)  \\ 
			& $\theta=2.5$ & 3.7349 & 9.2319 & 1.2349 & (1.094, 10.00)  \\ 
			& $\delta=0.25$ & 0.2529 & 0.0696 & 0.0029 & (0.010, 0.796)  \\
			\hline
			\multirow{4}{1em}{150} & $\alpha=1.5$ & 1.6100 & 0.1246 & 0.1100 & (1.009, 2.441) \\
			& $\beta=2.0$ & 2.0247& 0.0572 & 0.0247 & (1.636, 2.522)  \\ 
			& $\theta=2.0$ & 3.3929 & 5.8393 & 0.8929 & (1.196, 10.00) \\ 
			& $\delta=0.25$ & 0.2397 & 0.0610 & -0.0103 & (0.010, 0.765)  \\
			\hline
			\multirow{4}{1em}{250} & $\alpha=1.5$ & 1.5910 & 0.0638 & 0.0910 & (1.130, 2.095) \\
			& $\beta=2.0$ & 2.0116 & 0.0329 & 0.0116 & (1.694, 2.410) \\ 
			& $\theta=2.5$ & 3.1178 & 3.6097 & 0.6178 & (1.398, 10.00)  \\ 
			& $\delta=0.25$ & 0.2244 & 0.0517 & -0.0256 & (0.010, 0.747)  \\
			\hline
			\multirow{4}{1em}{500} & $\alpha=1.5$ & 1.5818 & 0.0266 & 0.0818 & (1.265, 1.824)\\
			& $\beta=2.0$ & 2.0045 & 0.0157 & 0.0045 & (1.767, 2.250)  \\ 
			& $\theta=2.5$ & 2.7440 & 1.0034 & 0.2440 & (1.605, 5.414)  \\ 
			& $\delta=0.25$ & 0.1888 & 0.0367 & -0.0612 & (0.010, 0.619)  \\
			\hline
		\end{tabular}
\end{table}

 \begin{table}[ht]
	\centering
	\caption{Case-2: Simulation study for different parameters values and 95\% C.I.}
	\normalsize
		\begin{tabular}{ |c|c|c |c| c| c|} 
			\hline
			Sample Size & Parameters & MLE & MSEs(MLE) & Bias(MLE)  & 95\% C.I.  \\
			\hline
			\multirow{4}{1em}{25} & $\alpha=1.0$ & 1.3401 & 0.3772 & 0.3401 & (0.576, 2.438) \\
			& $\beta=3.0$ & 3.4652& 1.9334 & 0.4652 & (2.087, 7.634)  \\ 
			& $\theta=2.5$ & 5.5018 & 25.3541 & 3.0018 & (0.447, 10.00)  \\ 
			& $\delta=0.5$ & 0.2981 & 0.1101 & -0.2019 & (0.010, 0.834)  \\
			\hline
			
			\multirow{4}{1em}{50} & $\alpha=1.0$ & 1.4000 & 0.4192 & 0.400 & (0.717, 2.509) \\
			& $\beta=3.0$ & 3.1811& 0.5500 & 0.1811 & (2.235, 4.854)  \\ 
			& $\theta=2.5$ & 4.5963 & 16.9078 & 2.0978 & (0.708, 10.00)  \\ 
			& $\delta=0.5$ & 0.1601 & 0.1470 & -0.3399 & (0.010, 0.636)  \\
			\hline
			\multirow{4}{1em}{100} & $\alpha=1.0$ & 1.3089 & 0.2383 & 0.3089 & (0.847, 2.351) \\
			& $\beta=3.0$ & 3.0750 & 0.1845 & 0.0750 & (2.396, 4.021)  \\ 
			& $\theta=2.5$ & 3.7553 & 9.2352 & 1.2353 & (1.094, 10.00)  \\ 
			& $\delta=0.5$ & 0.1680 & 0.1361 & -0.3314 & (0.010, 0.616)  \\
			\hline
			\multirow{4}{1em}{150} & $\alpha=1.0$ & 1.2821 & 0.1782 & 0.2821 & (0.862, 2.224) \\
			& $\beta=3.0$ & 3.0370 & 0.1288 & 0.0370 & (2.454, 3.783)  \\ 
			& $\theta=2.5$ & 3.3932 & 5.8424 & 0.6181 & (1.196, 10.00) \\ 
			& $\delta=0.5$ & 0.1724 & 0.1284 & -0.3276 & (0.010, 0.537)  \\
			\hline
			\multirow{4}{1em}{250} & $\alpha=1.0$ & 1.2538 & 0.1244 & 0.2538 & (0.947, 1.924) \\
			& $\beta=3.0$ & 3.0174 & 0.0741 & 0.0174 & (2.541, 3.615) \\ 
			& $\theta=2.5$ & 3.1181 & 3.6126 & 0.6181 & (1.398, 10.00)  \\ 
			& $\delta=0.5$ & 0.1701 & 0.1259 & -0.3299 & (0.010, 0.496)  \\
			\hline
			\multirow{4}{1em}{500} & $\alpha=1.0$ & 1.2208 & 0.0707 & 0.2208 & (1.017, 1.581)\\
			& $\beta=3.0$ & 3.0067 & 0.0353 & 0.0067 & (2.650, 3.376)  \\ 
			& $\theta=2.5$ & 2.7443 & 1.0045 & 0.2443 & (1.605, 5.416)  \\ 
			& $\delta=0.5$ & 0.1533 & 0.1298 & -0.3467 & (0.010, 0.366)  \\
			\hline
		\end{tabular}
\end{table}

\begin{table}[ht]
	\centering
	\caption{Case-3 : Simulation study for different parameters values and 95\% C.I.}
	\normalsize
	\begin{tabular}{ |c|c|c |c| c| c|} 
		\hline
		Sample Size & Parameters & MLE & MSEs(MLE) & Bias(MLE)  & 95\% C.I.  \\
		\hline
		\multirow{4}{1em}{25} & $\alpha=1.5$ & 2.3634 & 1.2824 & 0.8634 & (1.218, 3.921) \\
		& $\beta=3.5$ & 4.0230& 2.6153 & 0.5230 & (2.348, 9.648)  \\ 
		& $\theta=2.0$ & 4.9300 & 24.9229 & 2.9300 & (0.371, 10.00)  \\ 
		& $\delta=0.75$ & 0.2592 & 0.3374 & -0.4908 & (0.010, 0.877)  \\
		\hline
		
		\multirow{4}{1em}{50} & $\alpha=1.5$ & 2.3158 & 1.0457 & 0.8158 & (1.390, 3.834) \\
		& $\beta=3.5$ & 3.7059 & 0.8253 & 0.2059 & (2.563, 5.845)  \\ 
		& $\theta=2.0$ & 3.8026 & 14.2176 & 1.8026 & (0.609, 10.00)  \\ 
		& $\delta=0.75$ & 0.1959 & 0.3770 & -0.5541 & (0.010, 0.830)  \\
		\hline
		\multirow{4}{1em}{100} & $\alpha=1.5$ & 2.2435 & 0.7369 & 0.7435 & (1.614, 3.323) \\
		& $\beta=3.5$ & 3.5860 & 0.2724 & 0.0860 & (2.773, 4.710)  \\ 
		& $\theta=2.0$ & 2.9316 & 6.6118 & 0.9316 & (0.926, 10.00)  \\ 
		& $\delta=0.75$ & 0.1453 & 0.4177 & -0.6047 & (0.010, 0.770)  \\
		\hline
		\multirow{4}{1em}{150} & $\alpha=1.5$ & 2.2317 & 0.6360 & 0.7317 & (1.664, 2.939) \\
		& $\beta=3.5$ & 3.5448 & 0.1876 & 0.0448 & (2.831, 4.457)  \\ 
		& $\theta=2.0$ & 3.5604 & 3.0849 & 0.5604 & (1.010, 8.467) \\ 
		& $\delta=0.75$ & 0.1208 & 0.4341 & -0.6292 & (0.010, 0.680)  \\
		\hline
		\multirow{4}{1em}{250} & $\alpha=1.5$ & 1.2329 & 0.6025 & 0.7329 & (1.763, 2.703) \\
		& $\beta=3.5$ & 3.5219 & 0.1071 & 0.0219 & (2.942, 4.234) \\ 
		& $\theta=2.0$ & 2.3579 & 1.6197 & 0.3579 & (1.186, 5.620)  \\ 
		& $\delta=0.75$ & 0.0898 & 0.4612 & -0.6602 & (0.010, 0.570)  \\
		\hline
		\multirow{4}{1em}{500} & $\alpha=1.5$ & 0.2225 & 0.5537 & 0.7225 & (1.883, 2.499)\\
		& $\beta=3.5$ & 3.5098 & 0.0502& 0.0098 & (3.094, 3.956)  \\ 
		& $\theta=2.0$ & 2.1281 & 0.3619 & 0.1281 & (1.357, 3.671)  \\ 
		& $\delta=0.75$ & 0.0529 & 0.4959 & -0.6971 & (0.010, 0.407)  \\
		\hline
	\end{tabular}
\end{table}
\newpage
\section{Application}
This section applies the new generalized-Fisk model to the failure times data and compares it with some of its rival models.  Chipepa et al. (2020) stated failure times in hours for 101 participants as the source of the data used in this section's study. A further description of the data can be found at [(Barlow et al. 1984), (Andrews and Herzberg, 2012)].  Based on Moss (2004), the introduction section of our study highlights some of the characteristics of the practical data.  The information is available in the appendix under the term dataFT for section (6.2).  The box plot of the data is shown in Figure 2, and Table 4 provides descriptive summary measures of the data.

\begin{table}[ht]
	\centering
	\caption{Results of the descriptive summary for the Failure Times data in hours}
	\normalsize
	\begin{tabular}{ c c c c c c } 
		\hline\hline
		Minimum & $1^{st}$ Qu. & Median & Mean & $3^{rd}$ Qu. & Maximum  \\
		\hline\hline
		0.010  & 0.240  & 0.800 &  1.025  & 1.450 &  7.890\\
		\hline
\end{tabular}
\end{table}	

\begin{figure}[htp]\label{fig2}
	\centering
	\includegraphics[width=9cm]{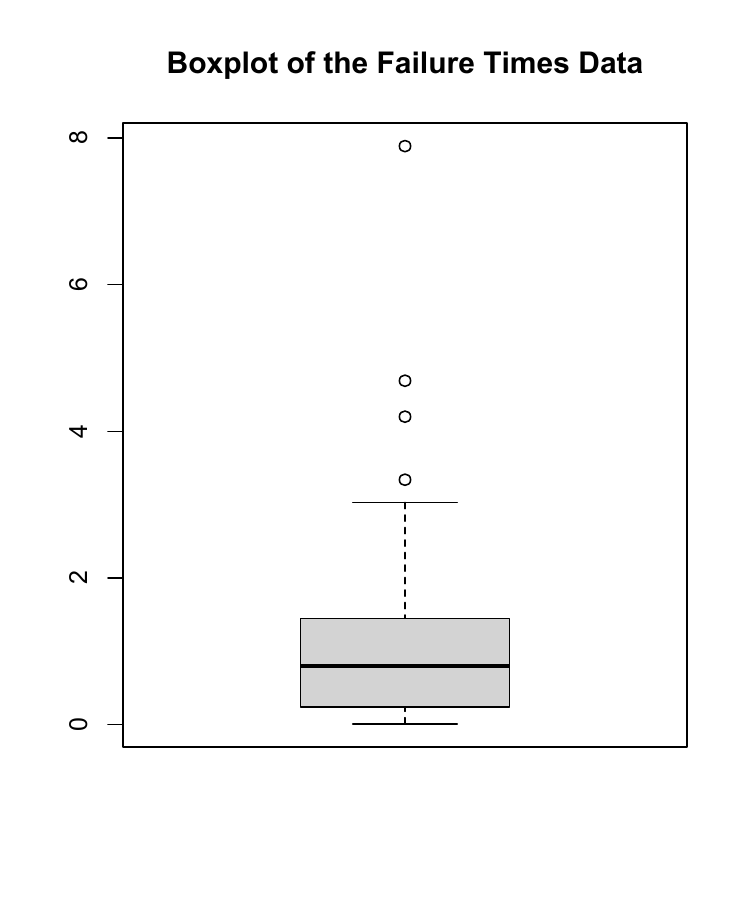}
	\caption{A box-plot graph showing the failure times in hours data}
	
\end{figure}
\newpage
The six descriptive summary measures of the data are provided in Table 4 and Figure 2.  With a range of 7.88 hours, the minimum and highest observations are 0.010 and 7.890 hours, respectively.  It appears that the data has been displayed to show the outliers.  Nonetheless, the disruption scale is tiny (less than tens or 7.88 hours).  As a result, the analysis treats this finding as such, and its disruption is accepted.  In addition to the mean (1.025) and the second quantile or median (0.800), the first and third quantiles are shown.  By contrasting the new Generalized-Fisk distribution with the three-parameter new extended flexible Weibull (NEx-FW), four-parameter flexible Weibull (FW), Four-parameter Kumaraswamy Weibull(Ku-W), and a three-parameter Zubair-Weibull (Z-W), and five-parameter Kumaraswamy Weibull Poisson (KWP), we demonstrate how well it fits the data. The CDFs of the competing models are:
\begin{itemize}
	\item 1. New Extended flexible Weibull (NEx-FW):
	\begin{equation*}
		G(x;\gamma, \beta, \theta) = 1-e^{-e^{\beta x^{\gamma}+\theta x^2}}; ~~~ x, \gamma, \beta, \theta > 0.
	\end{equation*}
	\item 2. Flexible Weibull(FW):
	\begin{equation*}
	G(x; \alpha, \gamma, \beta, \theta) = 1-e^{-e^{\beta x^{\gamma}+\theta x^{\alpha}}}; ~~~ x, \gamma, \alpha, \beta, \theta > 0.
	\end{equation*}
	\item 3. Kumaraswamy Weibull Poisson (KWP):
	\begin{equation*}
		G(x; a, b, c, \beta, \lambda) = \frac{1-e^{-\lambda[1-(1-(1-e^{-(\beta x)^c})^a)^b]}}{1-e^{-\lambda}}; ~~~ x, a, b, c, \beta, \lambda >0.
	\end{equation*}
	\item 4. Kumaraswamy Weibull (Ku-W):
	\begin{equation*}
		G(x; a, b, \gamma, \theta) = 1-(1-(1-e^{-\gamma x^{\theta}})^a)^b; ~~~ x, a, b, \gamma, \theta >0.
	\end{equation*}
	\item 5 . Z-Weibull:
	\begin{equation*}
		G(x; \alpha, \gamma, \theta) = \frac{e^{\alpha(1-e^{-\gamma x^{\theta}})}-1}{e^{(\alpha)}-1}; ~~~ \alpha, \gamma, \theta > 0, x\geq 0. 
	\end{equation*}
\end{itemize} 
The following model adequacy measurements, also known as goodness-of-fit tests, are used for model comparison.  These include the Cramer-von Mises (CM) statistic, the Hannan-Quinn Information Criterion (HQIC), the Bayesian Information Criterion (BIC), the consistent AIC (CAIC), and the AIC.  Thus, the model that best fits the data is the one that has the lowest value for these metrics.  Accordingly, the findings shown in Table (6) demonstrate that our suggested model, A new Generalized-Fisk, performs better for discrimination than the five competing models based on this comparison. The results for MLEs and Standard Errors (SEs) and model adequacy measures for the fitted models are summarized in Tables (5) and (6), respectively.  It is easy to see from these tables that the new generalized-Fisk model has improved, and its performance shows that it is the best model among the available options.

\begin{sidewaystable}
	\centering
	\caption{\textbf{MLE estimates of the parameters and the corresponding Standard Errors for the fitted models.}}
	\normalsize
	\begin{tabular}{ c c c c c c c c c c} 
		\hline\hline
		Distribution & $\hat{\alpha}$ & $\hat{\beta}$ & $\hat{\gamma}$ & $\hat{\delta}$ & $\hat{\theta}$ & $\hat{\lambda}$ & $\hat{a}$ & $\hat{b}$ & $\hat{c}$  \\
		\hline\hline
	\multirow{2}{5em}{NG-F}  & 5.991  & 0.982 &  & 0.373  &  10.00 & & & & \\
	& (2.657) & (0.012) & &  (0.074) & (4.904) & & & \\
		\hline
	\multirow{2}{5em}{NEx-FW} & &	0.457 & 0.648 &  & 0.011 & \\
	& & (0.044) & (0.062) & & (0.005) & \\
	\hline
	\multirow{2}{5em}{FW} & 0.630 & 0.297 & 0.626 & & 0.355 & \\
	& (0.178) & ($1.3\times 10^4$) & (0.216) & & ($1.3\times 10^4$) & \\
	\hline
	\multirow{2}{5em}{Ku-W} & 3.313 & & 3.313 & & 1.032 & & 0.734 & 0.270 & \\
	& (0.149) & & (0.149) & & (0.100) & & (0.172) & (0.035) & \\
	\hline
	\multirow{2}{5em}{Z-W} & 1.539 & &2.177 & & 0.558 & & & \\
	& (0.911) & & (0.329) & & (0.066) & & &\\
	\hline
	\multirow{2}{5em}{KWP} & & 0.107 & & & & 6.652 & 0.211 & 1.300 & 4.509\\
	& & (0.033) & &&  & (3.686) & (0.023) & (1.002) & (0.035) \\
	\hline
	\end{tabular}
\end{sidewaystable}	

\begin{table}[ht]
	\centering
	\caption{\textbf{Measures of model appropriateness for the fitted model.}}
	\normalsize
	\begin{tabular}{ c c c c c c c c c c} 
		\hline\hline
		Distribution & AIC & BIC & C AIC & HQIC & CM & -2logL \\
		\hline\hline
		NG-F & 210.65 & 221.54 & 211.43 & 214.65 & 0.09 & 101.32 \\
		\hline
		Ku-W & 213.25 & 223.71 & 213.67 & 217.65 & 0.15 & 103.54 \\
	\hline
	    Z-W & 214.45 & 224.78 & 215.65 & 218.39 & 0.27 & 105.67 \\
	    \hline
	    KWP & 216.76 & 229.54 & 219.24 & 220.19 & 0.24 & 106.86 \\
	    \hline
	    FW & 422.65 & 430.93 & 421.65 & 426.71 & 0.38 & 208.39 \\
	    \hline
	    NEx-FW & 515.42 & 523.12 & 516.29 & 517.36 & 0.85 & 256.38 \\
	    \hline
	\end{tabular}
\end{table}	
\newpage
\section{Conclusion}
The introduction of a new family of distributions aims to support the use of biomedical engineering, lifespans, survival analysis, and health.  A detailed discussion of the suggested family is provided for the NG-Fisk sub-family of distributions, which is created by using the ordinary log-logistic distribution as the baseline distribution in the NG-X family of distributions.  A detailed discussion is given of a few fundamental statistical features. By using the model discrimination techniques and five common distributions as competitors, the performance of this unique family member is evaluated.  The evaluation analysis demonstrates that the suggested distribution performs better than the other five alternatives.  According to the simulation study, as the sample size grows, the estimated parameter values remain relatively constant and approach the genuine parameter values.  It is demonstrated to be more useful and preferred than counterset models in biomedical engineering, survival analysis, and longevity.
Future research in the same field will focus on multivariate extension.  Researchers in the same field will have to cope with overdispersion and correlation by conducting simulation studies and evaluating the importance of the regression model parameters for frequently measured over-dispersed time-to-event data.

\section*{Data availability statement}
The appendix contains the first set of data, and upon request, the relevant author can provide the second set.

\section*{Conflicts of interest}
The authors state that there is no any form of conflict of interest affecting the publishing of this article.

\section*{Acknowledgement(s)}
I thank Ministry of Tribal Affairs-National Fellowship Scheme For Higher Education of ST Students(NFST) for providing me Junior Research Fellowship (JRF) and Senior Research Fellowship (SRF) (award no: 201819-NFST-TEL-00347).

\section*{Funding statement}
This study is supported by the Banaras Hindu University, Varanasi.

\section*{Appendix}
Data for Section 6.2\\
dataFT=c(4.69, 0.01, 1.51, 0.02, 7.89, 0.03, 1.11, 0.04, 0.05, 0.06, 0.07, 0.07, 0.08, 0.09, 0.09, 0.10, 0.10, 0.11, 0.11, 0.12, 0.13, 0.18, 0.19, 0.20, 0.23, 0.24, 0.24, 0.29, 0.34, 0.35, 0.36, 0.38, 0.40, 0.42, 0.43, 0.52, 0.54, 0.56, 0.60, 0.60, 0.63, 0.65, 0.67, 0.68, 0.72, 0.72, 0.72, 0.73, 0.79, 0.79, 0.80, 0.80, 0.83, 0.85, 0.90, 0.92, 0.95, 0.99, 1.00, 1.01, 1.02, 1.03, 1.05, 1.10, 1.10, 0.03, 1.15, 1.18, 1.20, 1.29, 1.31, 1.33, 1.34, 1.40, 1.43, 1.45, 1.50, 0.02, 1.52, 1.53, 1.54, 1.54, 1.55, 1.58, 1.60, 1.63, 1.64, 1.80, 1.80, 1.81, 2.02, 2.05, 2.14, 2.17, 2.33, 3.03, 3.03, 3.34, 4.20, 0.01, 0.02).

\end{document}